%% file: MNmodels.v2.tex
\documentclass[useAMS,usenatbib,usegraphicx]{mn2e}

\usepackage{graphicx}
\usepackage[latin1]{inputenc}
\usepackage{color}
\usepackage{times}
\usepackage{natbib}
\usepackage{setspace}
\newif\ifAMStwofonts
\AMStwofontstrue
\definecolor{red}{rgb}{1,0.,0.}

\newcommand{\munich}{{\sc l-galaxies }}
\newcommand{\lcdm}{$\Lambda$CDM }
\newcommand{\masneu}{\sum_i m_{\nu_i} }
\newcommand{\msun}{{\rm M}_\odot}

\def\lesssim{\lower.5ex\hbox{$\; \buildrel < \over \sim \;$}}
\def\gtrsim{\lower.5ex\hbox{$\; \buildrel > \over \sim \;$}}
\title[Galaxy formation and massive neutrinos] {Semi-Analytic Galaxy
  Formation in Massive Neutrino Cosmologies}

\author[Fontanot et al.]{
  \parbox[t]{\textwidth}{Fabio Fontanot$^1$\thanks{E-mail:
      fontanot@oats.inaf.it}, Francisco Villaescusa-Navarro$^{1,2}$,
    Davide Bianchi$^{3}$, Matteo Viel$^{1,2}$}
    \vspace*{8pt}\\
    $^1$ INAF - Astronomical Observatory of Trieste, via G.B. Tiepolo 11, I-34143 Trieste, Italy \\
    $^2$ INFN - Sezione di Trieste, Via Valerio 2, 34127 Trieste, Italy \\
    $^3$ INAF-Osservatorio Astronomico di Brera, Via Bianchi 46, I-23807 Merate (LC), Italy\\
}

\begin{document}
\date{Accepted ... Received ...}

\maketitle

\begin{abstract} 
The constraints on neutrino masses led to the revision of their
cosmological role, since the existence of a cosmological neutrino
background is a clear prediction of the standard cosmological
model. In this paper, we study the impact of such background on the
spatial distribution of both Dark Matter (DM) and galaxies, by
coupling $N$-body numerical simulations with semi-analytic models
(SAMs) of galaxy formation. Cosmological simulations including massive
neutrinos predict a slower evolution of DM perturbations with respect
to the \lcdm runs with the same initial conditions and a suppression
on the matter power spectrum on small and intermediate scales, thus
impacting on the predicted properties of galaxy populations. We
explicitly show that most of these deviations are driven by the
different $\sigma_8$ predicted for cosmologies including a massive
neutrino background. We conclude that independent estimates of
$\sigma_8$ are needed, in order to unambiguously characterise the
effect of this background on the growth of structures. Galaxy
properties alone are a weak tracer of deviations with respect to the
\lcdm run, but their combination with the overall matter distribution
at all scales allows to disentangle between different cosmological
models. Moreover, these deviations go on opposite direction with
respect to competing models like modified gravity, thus weakening any
detectable cosmological signal. Given the ubiquitous presence of a
neutrino background, these effects have to be taken into account in
future missions aimed at constraining the properties of the ``Dark''
components of the Universe.
\end{abstract}


\section{Introduction}\label{sec:intro}
The accurate measurement of the value of cosmological parameters from
the Cosmic Microwave Background \citep{Planck_cosmpar,Hinshaw13}
opened a completely new window on the study of the basic properties of
our Universe. In particular, the role of the so-called ``Dark''
components, i.e. Dark Matter (DM) and Dark Energy (DE), as the main
contributors to the current energy density of the Universe has rised
considerable debate. Despite the undisputed successes of the standard
\lcdm cosmological scenario at physical scales ranging from the
Galactic to the large scale structure (LSS hereafter), the still
unknown properties of the ``Dark'' components remain a challenge to
our understanding of the Universe as a whole.

Numerous scenarios have been proposed in an attempt to explain the
origin and rise of such components: as an example, generalised DE
models overcome the theoretical difficulties related to the simplest
scenario based on a cosmological constant $\Lambda$ \citep[see
  e.g][for a review of the different DE
  scenarios]{Amendola13}. Besides, different models have been
suggested to explain the nature of the DM particle, based on some
assumptions on its phase-space density and/or interaction properties,
including extensions of the Particles Standard Model (e.g. axions,
supersymmetric or weakly interacting massive particles). Present
constraints based on cosmic microwave background (CMB) measurements
are in better agreement with $\Omega_{\rm m}$ being dominated by a
non-baryonic ``Cold'' DM particle (i.e. characterised by
non-relativistic velocities). It is worth stressing that, as long as
the ``cold'' component dominates, a small contribution from a
different DM species, with ``hotter'' properties, is possible. In
these mixed or ``Warm'' DM scenarios \citep[see e.g.][and references
  herein]{Maccio13, Viel13}, the growth of structures in both the
linear and non-linear regime is affected by the hottest component, due
to its relatively large free-streaming scale. Conversely, the
evolution of the LSS of the Universe provides strong constraints on
the maximum contribution of these hot species to the total DM budget.

The standard cosmological Big Bang theory predicts the existence of a
neutrino background \citep[see e.g.][]{LesgourguesPastor06} and
neutrinos contribute to the total radiation energy density in the
Early Universe, thus affecting the early nucleosynthesis of light
elements. Commonly considered as massless particles, the cosmological
role of neutrinos as DM candidates has been revived by the discovery
of the neutrino oscillation phenomenon
\citep{Cleveland98,Fogli12,Forero12}, which proved that at least two
of the three neutrino families should have a mass. It is worth
stressing that these experiments only provide information on the mass
square difference between the different neutrino families, which is
then converted into constraints on their total mass. In addition, CMB
experiments and galaxy surveys studied the shape of the matter power
spectrum and were able to put upper limits on the total neutrino mass
of the order of $\masneu<0.3$ eV \citep[see e.g.][and references
  herein]{Xia12, RiemerSorensen14, Costanzi14}. At variance with Warm
Dark Matter cosmologies, which can be viewed as ``exotic'' models
meant to solve a number of inconsistencies in the standard
cosmological model (like those related to halo profiles and subhalo
abundances), massive neutrinos are nowadays regarded as a fundamental
element in cosmology and constraining their masses is a key target in
order to explore physics beyond the standard model.

This paper is the third of a series aimed at the study of the
properties of galaxy populations as predicted by semi-analytic models
(SAMs) of galaxy formation and evolution in non-\lcdm cosmologies. In
the first two papers, we consider Early Dark Energy \citep[][hereafter
  Paper I]{Fontanot12c} and $f(r)$-Gravity \citep[][hereafter Paper
  II]{Fontanot13b} cosmologies and we discuss which observables are
the most suitable to distinguish these scenarios from a standard \lcdm
universe. In this paper, we expand this suite of mock galaxy
catalogues coupling SAMs with numerical simulations of massive
neutrino cosmologies. In the SAM framework (and in hydrodynamical
simulations as well), the relevant physical mechanisms acting on the
baryonic component and responsible for galaxy formation and evolution
(gas cooling, star formation, black hole accretion, feedbacks) are
modelled using simplified analytic prescriptions, which describe the
main dependencies, as a function of the physical properties of model
galaxies (stellar, gas and metal content, morphology), environment
(parent halo mass) and hierarchy (central or satellite). Such models
are thus characterised by a number of free parameters, usually
calibrated against a well defined set of low-redshift
observations. This approach is flexible enough to test different
prescriptions for the relevant processes and their interplay, thus
providing key insight in our understanding of the complex processes
leading to the built up of the different galaxy populations. However,
a number of tensions between model predictions and observational
constraints are still present \citep[see e.g][among
  others]{McCarthy07, BoylanKolchin12, Weinmann12, Henriques13,
  Wilman13} pointing to the need for a revision of some key
ingredients. Moreover, the SAM approach implies a relevant level of
intrinsic degeneracy among the different parameters
\citep{Henriques09}, which is exacerbated by the fact that different
groups made different choices for the (equally plausible) modelling of
the main processes. The predictions of independently developed SAMs
show a reasonable agreement for a number of key quantities \citep[see
  e.g.][]{Fontanot09b, Fontanot12a}. Nonetheless it is of fundamental
importance, in the context of future space missions aimed at a better
characterisation of DE and DM \citep[like the EUCLID
  mission][]{Laureijs11}, to identify modifications of galaxy
properties that can be uniquely associated with the different
cosmological frameworks and define suitable statistical tests based on
galaxy populations able to disentangle such models from the standard
cosmological model. In fact, most of the key cosmological probes
proposed in the context of such missions, ultimately rely on the
spatial distribution of galaxy populations, used as tracers of the
underlying LSS at different redshifts.

This paper is organised as follows. In Section~\ref{sec:models}, we
introduce the cosmological numerical simulations and semi-analytic
models we use in our analysis. We then present the predicted galaxy
properties and compare them among different cosmologies in
Section~\ref{sec:results}. Finally, we discuss our conclusions in
Section~\ref{sec:final}.

\section{Models}\label{sec:models}
\begin{table*}
  \caption{Cosmological Parameters for the cosmological simulation
    suite.}
  \label{tab:suite}
  \renewcommand{\footnoterule}{} \centering
  \begin{tabular}{ccccccccc}
    \hline
     & $\Omega_{\Lambda}$ & $\masneu$ & $\Omega_{\nu}$ & $\Omega_m$ & Resolution [$h^{-1} \msun$] & $h$ & $\sigma_8$ \\
    \hline
    \lcdm & 0.6825 & 0.0 eV & 0.0    & 0.3175 & 6.57 $\times 10^8$ & 0.6711 & 0.834 \\
    NU03  & 0.6825 & 0.3 eV & 0.0072 & 0.3175 & 6.42 $\times 10^8$ & 0.6711 & 0.763  \\
    NU06  & 0.6825 & 0.6 eV & 0.0143 & 0.3175 & 6.27 $\times 10^8$ & 0.6711 & 0.692  \\
    \hline                            
    N3s8  & 0.6825 & 0.3 eV & 0.0072 & 0.3175 & 6.42 $\times 10^8$ & 0.6711 & 0.834 \\
    N6s8  & 0.6825 & 0.6 eV & 0.0143 & 0.3175 & 6.27 $\times 10^8$ & 0.6711 & 0.834 \\
    \hline 
  \end{tabular}
\end{table*}

\subsection{Massive neutrino cosmologies.}

Massive neutrinos affect the growth of cosmological LSS at different
scales. At the linear order, they shift the matter-radiation equality
time, stretching out the radiation-dominated epoch, while in the
matter-dominated era they slow down the growth of matter
perturbations. The combination of these two effects determines a
suppression of the matter power spectrum on small scales
\citep{LesgourguesPastor06}. In the fully non-linear regime, on the
other hand, massive neutrinos induce a variety of effects and in order
to properly characterise their impact on the matter power spectrum
N-body simulations have been used \citep[see e.g][]{Brandbyge08,
  Viel10, Agarwal11, Bird11, Wagner12}. Those works have pointed out
that the suppression of power is higher in the fully non-linear regime
than in linear theory. However, unlike the linear case, the
suppression is redshift- and scale-dependent.

At the decoupling time, the momentum distribution of the neutrinos is
expected to follow the Fermi-Dirac distribution: thus, a small
fraction of the cosmic neutrinos will have velocities low enough to
cluster within the dark matter halos \citep[see e.g.][and reference
  herein]{Brandbyge10, IchikiTakada12, LoVerde14a} and form, via
gravitational collapse, halos of neutrinos. These structures modify
the total matter density profile and in principle may be detected via
gravitational lensing in future surveys \citep{VillaescusaNavarro11}.

Recently, in a series of works \citep{VillaescusaNavarro14,
  Castorina13, Costanzi13b} it has been shown that massive neutrinos
induce a scale-dependent bias on large-scales. In particular
\cite{Castorina13} pointed out that the scale-dependence almost
disappears if the bias is defined as the ratio between the halo power
spectrum and the cold dark matter (CDM) power spectrum. Massive
neutrinos have also an impact on the halo mass function, as the same
authors have shown that in massive neutrino cosmologies this
constraint can be entirely, and universally, described in terms of the
properties of the CDM field alone \citep[see also][where this aspect
  has been suggested for the first time]{Brandbyge10}. Additional
signatures of the presence of a massive neutrino background are also
expected on the Ly-$\alpha$ forest \cite{Viel10,
  VillaescusaNavarro13b, Rossi14}, on the Sunyaev-Zeldovich and X-ray
properties of galaxy clusters \citep{Roncarelli14}, on the
redshift-space distortions \cite{Marulli11}, and on the cosmic voids
statistics \cite{VillaescusaNavarro13b}.

The first attempt to populate dark matter halos with galaxies in
massive neutrino cosmologies has been carried out in
\cite{VillaescusaNavarro14} using a halo occupation distribution model
(HOD). In this paper, we plan to extend this study and we investigate
the distribution and properties of different galaxy populations, as
predicted by SAMs. It is well known that the effects induced by
massive neutrinos, through a non-vanishing value of $\Omega_\nu$, can
be mimicked by a standard cosmological model with a different
normalisation of the matter power spectrum: this is the so-called
$\Omega_\nu-\sigma_8$ degeneracy. In this paper we will also explore
whether this degeneracy can be broken by means of mock galaxy
catalogues.

\subsection{Numerical simulations}

In this paper, we consider a set of numerical simulations similar to
those used in \citet{VillaescusaNavarro14}, using a modified version
of the cosmological code {\sc gadget3}: in order to follow the
evolution of the LSS on non-linear scales, these runs employ the
so-called {\it particle method}, which explicitly incorporates
neutrinos in the simulations as particles. In all simulations, we
assume a flat universe consistent with Planck cosmology
\citep{Planck_cosmpar}, with matter density parameter
$\Omega_m=0.3175$, Hubble parameter $h=0.6711$ and with Gaussian
density fluctuations with a scale-invariant primordial power spectrum
with spectral index $n=0.9624$. We generate initial conditions for all
the simulations using a modified version of the {\sc n-genic} code: we
set them to have the same matter power spectrum at the last scattering
surface and we impose the same phases and mode amplitudes to force a
similar realisation of the large-scale structure and allow a proper
object-by-object comparison. Since $\Omega_m$ is kept constant in all
runs, simulations with larger $\Omega_\nu$ have smaller $\Omega_{\rm
  cdm}$. We thus run a set of simulations with varying total neutrino
masses including $\masneu=0.0$ eV (i.e. a standard vanilla \lcdm),
$0.3$ eV (NU03), $0.6$ eV (NU06). In this reference simulation suite,
the normalisation of the power spectrum at early times is the same as
in a \lcdm universe with $\sigma_8=$ 0.834; nonetheless, the presence
of a massive neutrino component changes the linear structure growth as
a function of redshift and scale, and the actual $\sigma_8$, measured
at $z=0$ will be different from the corresponding \lcdm value. In an
attempt to break the degeneracy between $\masneu$ and $\sigma_8$ we
also run additional simulations with $\masneu=0.3$ eV and $0.6$ eV
where we vary the amplitude of the initial fluctuations to obtain the
same $\sigma_8$ value at $z=0$ as in the \lcdm realisation (N3s8,
N6s8).

We set up our simulations of periodic boxes of 100 $h^{-1} {\rm Mpc}$ on
a side using $512^3$ CDM and $512^3$ neutrino particles, corresponding
to a mass resolution of 6.57 $\times 10^8 h^{-1} \msun$ for the \lcdm
realisation (and slightly lower for the other runs, see
Table~\ref{tab:suite}). For each run, 63 simulations snapshots were
stored at the same redshifts used in the Millennium simulation
\citep{Springel05} and in Paper I and II. Group catalogues have been
constructed using a Friend-of-Friend algorithm with a linking length
of $0.2$ (in mean particle separation units), and gravitationally
bound substructures have been defined using {\sc subfind}
\citep{Springel01} (only subhalos that retain at least 32 particles
after the gravitational unbinding procedure were considered). We then
use the subhalo catalogues to define the merger tree histories as in
\citet{Springel05}.

\subsection{Semi-Analytic Models}
\begin{figure*}
  \centerline{ \includegraphics[width=18cm]{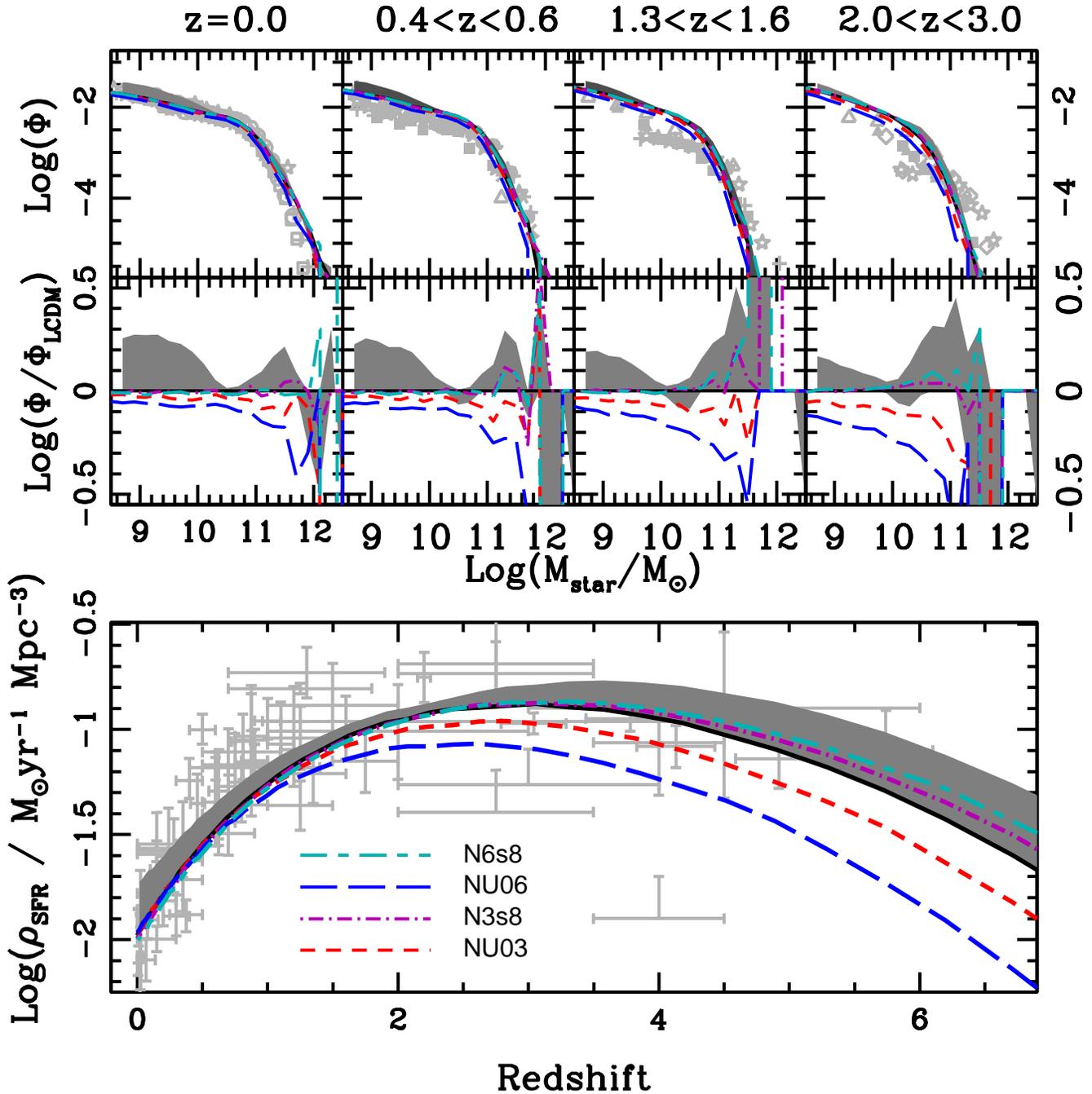} }
  \caption{SAM predictions in different massive neutrino cosmological
    scenarios for the redshift evolution of the predicted stellar mass
    function ({\it upper panel} - light grey points refer to the
    compilation from \citet{Fontanot09b}) and for the cosmic star
    formation rate density ({\it lower panel} - light grey points
    refer to the compilation from \citet{Hopkins04}). In each panel
    the solid black, long-dashed red, dot-dashed violet, short-dashed
    blue and short-long-dashed light blue lines refer to SAM
    predictions in \lcdm, NU03, N3s8, NU06 and N6s8 cosmologies,
    respectively, as labelled. Dark grey areas mark the distribution
    in the predictions between the \citet{Guo11},
    \citet{DeLuciaBlaizot07} and \citet{Croton06} SAMs for \lcdm
    cosmology.}\label{fig:gen_sam}
\end{figure*}
\begin{figure*}
  \centerline{ 
    \includegraphics[width=9cm]{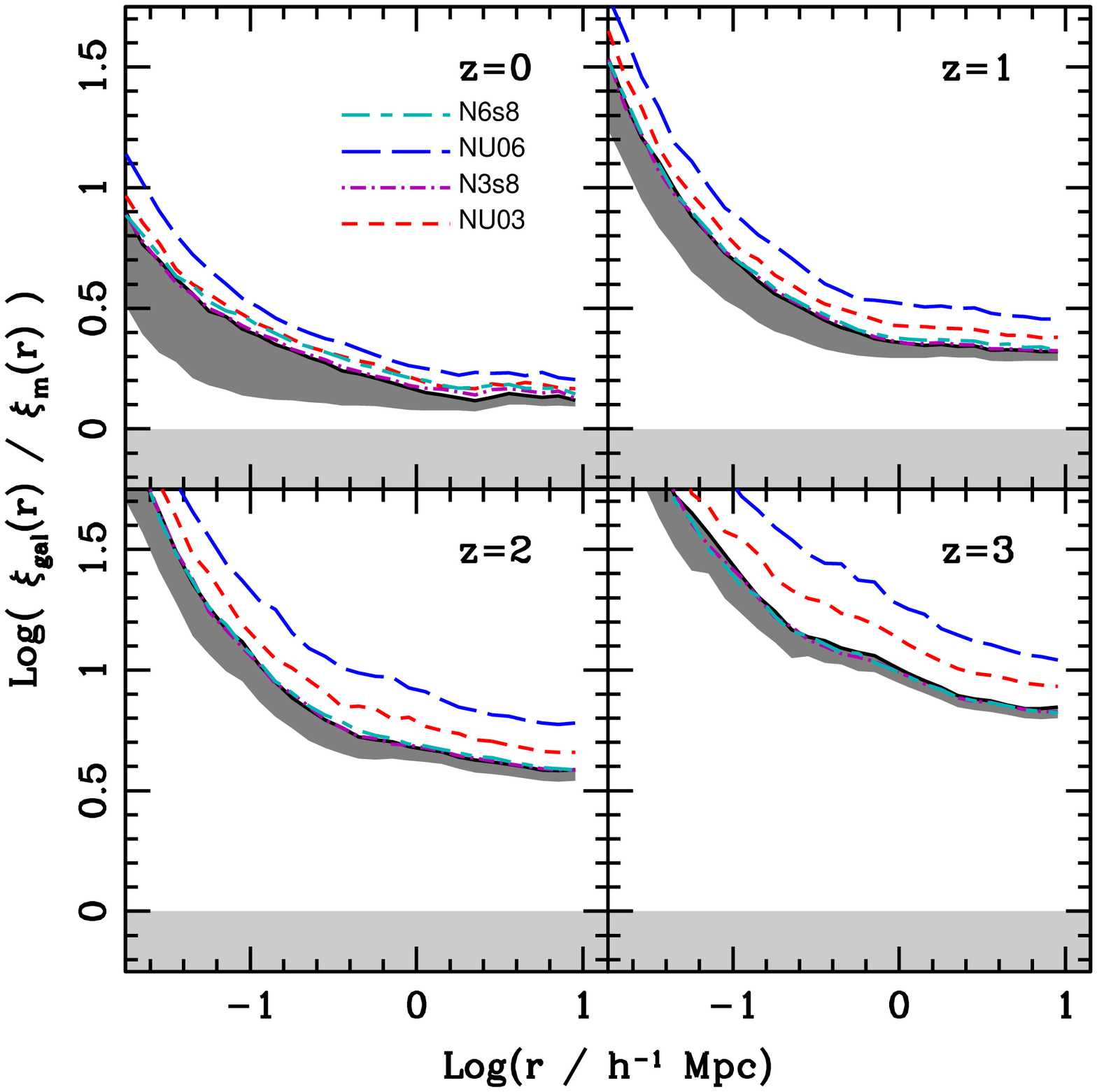} 
    \includegraphics[width=9cm]{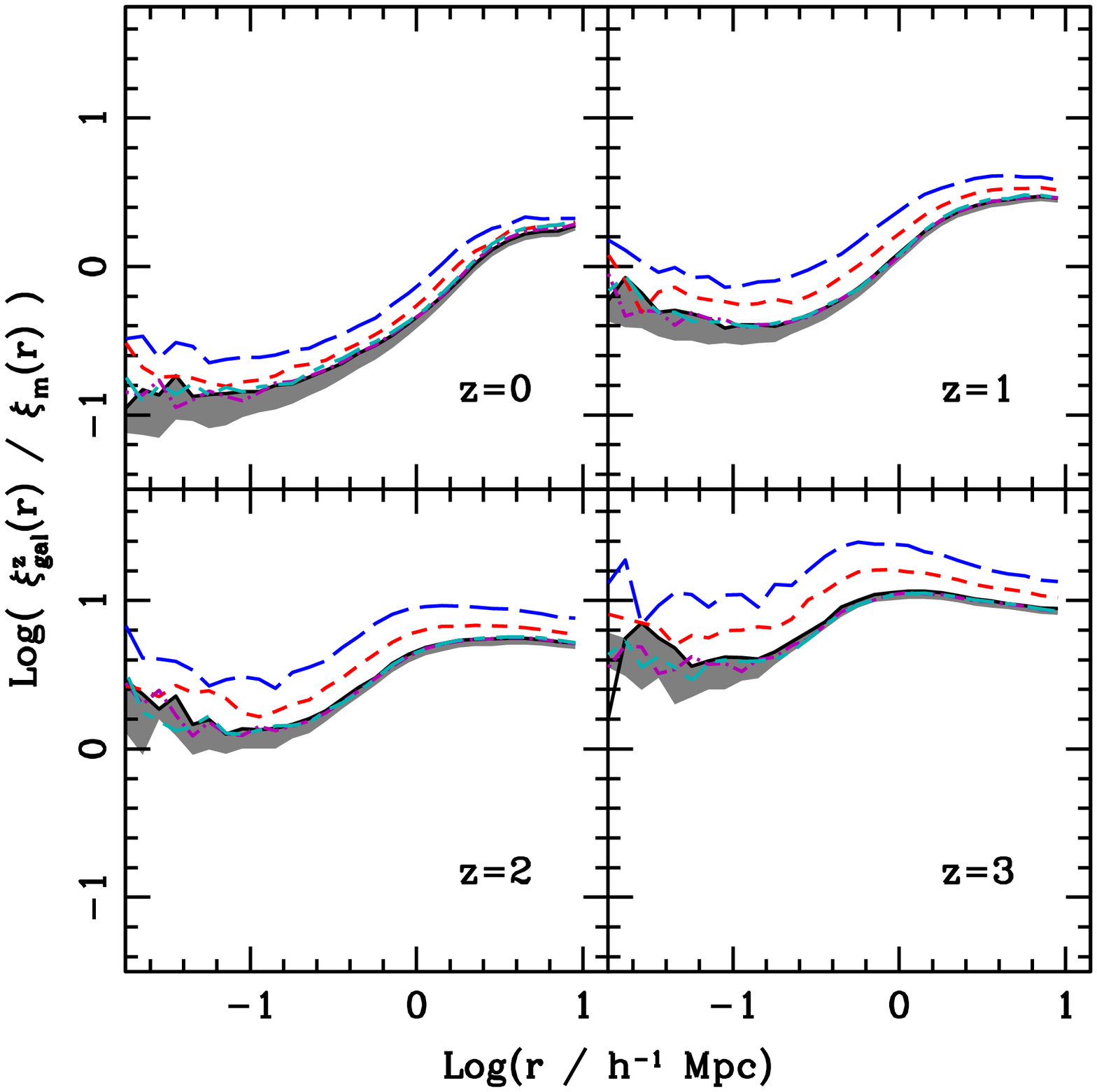} }
  \caption{Redshift evolution of galaxy bias in real ({\it left
      panel}) and redshift space ({\it right panel}) for different
    massive neutrino cosmologies. In each panel, only model galaxies
    with $M_\star > 10^9 \msun$ have been considered while computing
    the galaxy 2-points correlation functions. Models are labelled
    with the same line types, colours and shades as in
    Figure~\ref{fig:gen_sam}.}\label{fig:bias}
\end{figure*}
\begin{figure*}
  \centerline{ \includegraphics[width=18cm]{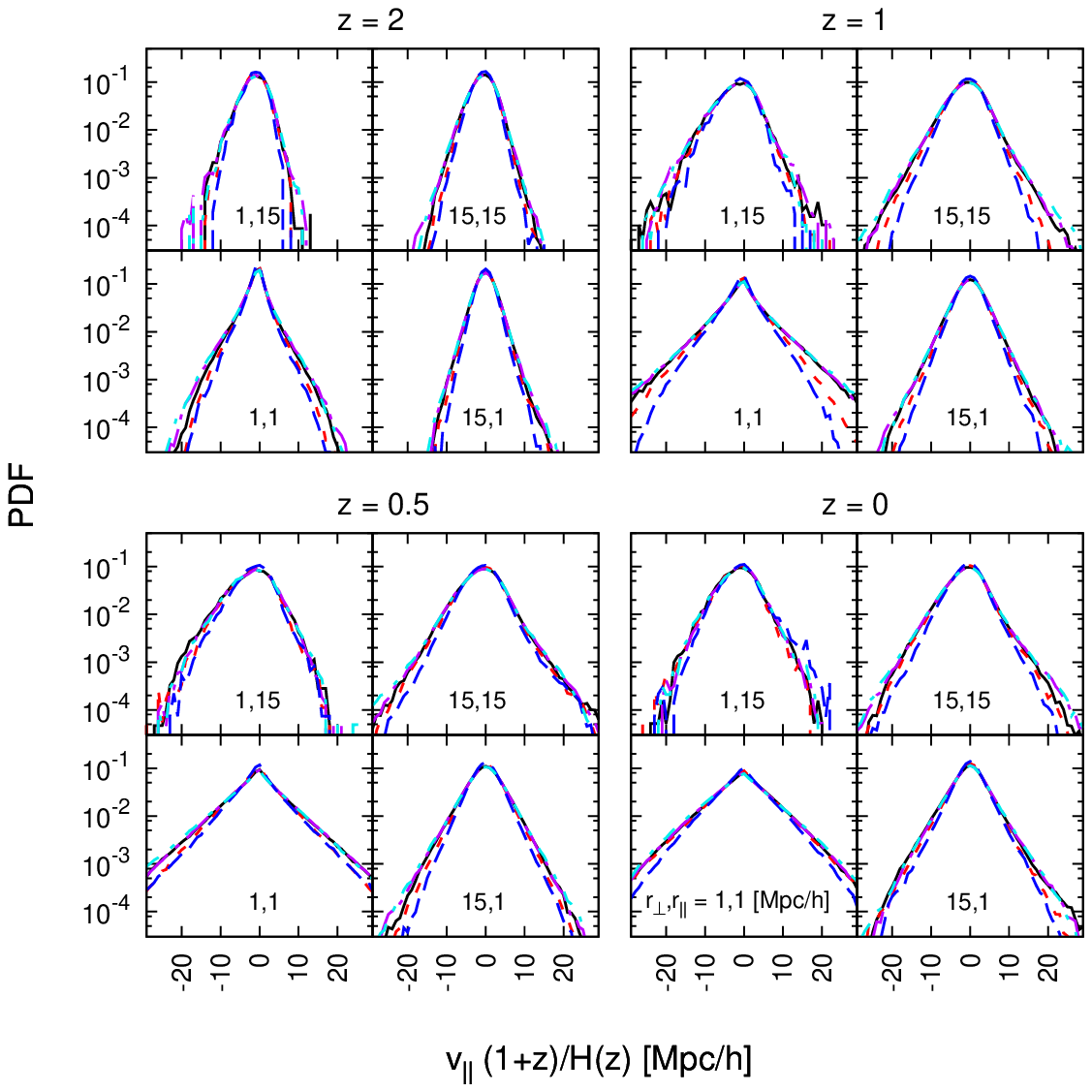} }
\caption{Pairwise galaxy velocity distribution along the line of sight
  for the massive neutrino cosmologies at four different
  redshifts. Velocities have been rescaled to comoving distances using
  the conformal Hubble function $\mathcal{H}=aH$. Each panel
  represents a different combination for values of the galaxy
  separation ($r_\perp$,$r_\parallel$), perpendicular and parallel to
  the line of sight, respectively (as labelled). Different
  cosmological models are marked by different linetypes and colour as
  in Figure~\ref{fig:gen_sam}.}\label{fig:pairwise}
\end{figure*}

In this work, we use the same approach we used in the previous papers
of this series. We consider three different versions of the \munich
semi-analytic model, based on the code originally developed by
\citet{Springel05}: those are, in historical order, the versions
described in \citet{Croton06}, \citet{DeLuciaBlaizot07} and
\citet{Guo11}. All these models share a common code structure and are
designed to run on the same merger tree histories defined in the
previous section. Moreover, they provide a representative set of
models characterised by different choices in the modelling of the
relevant galaxy formation physics\footnote{From the \citet{Croton06}
  to the \citet{DeLuciaBlaizot07} version, the main differences lie in
  the treatment of dynamical friction and merger times, the initial
  mass function (from Salpeter to Chabrier) and the dust modelling;
  from the \citet{DeLuciaBlaizot07} to the \citet{Guo11} version, the
  main changes involve the modelling of supernovae feedback, the
  treatment of satellite galaxy evolution, tidal stripping and
  mergers. In the following, the predictions of the \citet{Croton06}
  model have been converted to a Chabrier IMF by applying a constant
  shift (0.25 dex in stellar mass and 0.176 dex in star formation
  rate) to the original, Salpeter IMF calibrated, predictions.}, which
typically require a general re-calibration of the main model
parameters, against comparable sets of low-redshift reference
observations. Therefore, when these models are applied to the same
\lcdm cosmological simulation, we expect the scatter in their
predictions to be representative of the variance of SAM
predictions\footnote{We note that, since all models we consider use
  Millennium-like merger trees, we get rid of any additional source of
  noise due to the different merger tree formats used in the SAM
  framework, see e.g. \citet{Knebe14}.}.

Our reference version of the model is the same as proposed in the
original \citet{Guo11} paper: the presence of a massive neutrino
component affects mainly the growth of LSS (at variance with Paper I
and II we do not expect any effect neither in the Hubble Function nor
in the baryonic physics) and this information is completely defined in
the different merger tree histories. We will thus focus mainly in
understanding the effect of massive neutrino scenarios on galaxy
properties (like the assembly of stellar mass, the cosmic star
formation rate and galaxy clustering).

As in Paper I and II, we do not consider possible re-calibrations of
the \citet{Guo11} model and we rather prefer to keep the original
parameter set: this choice allows a direct comparison to the published
models and to highlight differences induced by changes in the
cosmology alone. This implies that the models with an increasingly
large contribution of neutrinos are not necessarily tuned to perform
best, as in the \lcdm case.

\section{Results}\label{sec:results}

As in Paper I and II, we compare the redshift evolution of selected
statistical properties of mock galaxy populations in the different
cosmologies. In particular, we consider the galaxy stellar mass
function (Figure~\ref{fig:gen_sam}, upper panel), the cosmic star
formation rate (Figure~\ref{fig:gen_sam}, lower panel), the galaxy
bias (both in real and redshift space, Figure~\ref{fig:bias}), and the
pairwise velocity distribution (Figure~\ref{fig:pairwise}). In the
following, only galaxies with $M_\star > 10^9 \msun$ have been
considered and in Figure~\ref{fig:gen_sam}, model predictions are
convolved with an estimate of the error associated with observational
constraints (i.e. a lognormal distribution with amplitude 0.25 and 0.3
for stellar masses and star formation rates, respectively). In all
figures, shaded areas represent the locus span by the predictions of
the three different SAMs when applied to the same \lcdm box, the black
solid line being the prediction of the \citet{Guo11} model: as we
discussed in the previous sections, we consider the shaded area as
representative of the variance between SAMs. The predictions of the
\citet{Guo11} model applied to massive neutrino realisations are
highlighted by different linetypes and colours: long dashed red,
dot-dashed violet, short dashed blue and long-short-dashed cyan refer
to the NU03, N3s8, NU06 and N6s8 runs, respectively.

Massive neutrino cosmologies induce systematic deviations in galaxy
properties with respect to \lcdm: in particular the slower growth of
structures is reflected in a smaller space density of galaxies at all
mass scales and redshifts, i.e. in a lower cosmic star formation
rate. The differences with respect to \lcdm are larger at the high
mass end of the stellar mass function, and tend to be small or
negligible at the low-mass end: this implies that the dwarf
overproduction problem \citep{Fontanot09b} is not reduced in massive
neutrino cosmologies. In realistic cases ($\masneu \simeq 0.3$ eV)
these deviations are of the same order of the intra-SAM variance, and
only models with relatively large values of $\masneu$ show relevant
deviations. Nonetheless, it is worth stressing that similar trends are
expected also for standard \lcdm realisations with different
$\sigma_8$ \citep[see e.g.][]{Wang08, Guo13}. Indeed, the results for
the N3s8 and N6s8 runs clearly show that most of the difference
between massive neutrino cosmologies and \lcdm are washed out, if we
force the former runs to have the same $\sigma_8$ at $z=0$: as a
consequence they became indistinguishable from a standard cosmological
model. Therefore, an independent estimate of $\sigma_8$ at different
redshifts is needed in order to use our results as constraints for
massive neutrino cosmologies.

In the left panel of Figure~\ref{fig:bias}, galaxy bias is estimated
from the ratio between the auto-correlation function of galaxies in
real space $\xi_{\rm gal}$ and the auto-correlation function of total
matter distribution in real space $\xi_{\rm m}$, using the
\citet{LandySzalay93} estimator. The latter quantity has been computed
combining the auto-correlation function of CDM and neutrino and the
cross-correlation among them (see Equation~12 in
\citealt{VillaescusaNavarro14}), and using a subsample of $10^6$ CDM
particles and $10^6$ neutrino particles randomly extracted from the
corresponding simulations. From the analysis of this plot we reach
similar conclusions with respect to Figure~\ref{fig:gen_sam}: massive
neutrino cosmologies show a clear increase in the bias at all scales
with respect to \lcdm model with the same power spectrum at
recombination. Moreover, it is possible to define a range of physical
scales and redshifts where the cosmological signal is clearly larger
than the variance between different SAM predictions. However, most of
these effects are connected to the different $\sigma_8$ evolution. The
same conclusions hold when galaxy bias is computed in redshift space
(using the ratio between the auto-correlation function for galaxies in
redshift space $\xi_{\rm gal}^{\rm z}$ and the auto-correlation
function of total matter distribution in real space,
Figure~\ref{fig:bias}, right panel), showing that the analysis in
redshift space disentangles different massive neutrino cosmologies as
efficiently as in the real space.

Finally, in Figure~\ref{fig:pairwise} we show the redshift evolution
of the pairwise galaxy velocity distribution along the line of sight
$\mathcal{P}(v_\parallel,r_\parallel,r_\perp)$, measured considering
fixed components of galaxy separation parallel ($r_\parallel$) and
perpendicular ($r_\perp$) to the line of sight \citep[see
  e.g.][]{Scoccimarro04}. The actual choice of reference separations
(1 and 15 ${\rm Mpc/h}$) has been motivated by the limited
cosmological volume considered in our boxes. Only galaxies with
$M_\star > 10^9 M_\odot$ have been considered and their
velocities\footnote{As in Paper II, we assume that the pairwise
  velocity is negative when galaxies are approaching each other and
  positive when they are receding.} have been rescaled using the
conformal Hubble function $\mathcal{H}=aH$ in order for the
distribution to represents the statistical displacement of galaxy
pairs from real to redshift space. The pairwise velocity distribution
is a reliable tracer of the anisotropy of redshift-space correlation
functions and the assembly and growth of LSS. As expected, given the
different growth history in massive neutrino cosmologies, there is
some statistical difference between SAM predictions relative to NU03
and NU06 and all other realisations: however, the effect is rather
small for realistic cases ($\masneu \sim 0.3$ eV).

By comparing Figure~\ref{fig:bias} and~\ref{fig:pairwise} with the
corresponding plots in Paper I and II, we noticed that the combined
deviations are clearly different from those predicted in the case of
Early Dark Energy or $f(r)$-gravity runs, showing that in principle it
should be possible to disentangle between these cosmologies using
these tests. Moreover, it is worth stressing that, since massive
neutrino cosmologies imprint opposite trends with respect to the
other models, the existence of a neutrino background (as predicted by
the standard cosmological model) has the net effect of smoothing any
signal coming from these. This has indeed been already pointed out by
simulations combining massive neutrinos and $f(r)$-gravity \citep[see
  e.g][]{Baldi14}.

\section{Conclusions \& Discussion}\label{sec:final}

In this paper we study the impact of cosmologies including massive
neutrinos on the properties of galaxy populations, as predicted by
SAMs. This work has important implications, since the existence of a
neutrino background (and its role in early nucleosynthesis) is a
robust prediction of the standard cosmological model, and given the
evidences in favour of massive neutrinos \citep{Cleveland98, Fogli12,
  Forero12}. In this paper, we couple a suite of $N$-body
CDM+neutrinos simulations with the \munich model \citep[in
  the][version]{Guo11}. We also consider earlier \munich versions to
constrain the variance in the SAM predictions when applied to the same
\lcdm realisation. Our results are compatible with our previous
findings (see Paper I and II) and similar studies based on coupling
Warm DM simulations with SAMs \citep{Kang13}: the presence of an
additional, but subdominant, hot/warm DM component leads to small but
systematic deviations in the global properties of galaxy
populations. It is worth stressing that most of the effects we find
are mainly driven by the lower $\sigma_8$ values predicted for
cosmological boxes including massive neutrinos with respect to \lcdm,
and due to the requirement that the amplitude of matter power spectrum
at recombination to be the same in all runs. Therefore, it is of
fundamental importance to have an independent and firm estimate of
$\sigma_8$ at different redshifts coming from other cosmological
constraints, in order to break the degeneracy. Given this estimate,
our results show the effects on structure formation of including
massive neutrinos in theoretical models of galaxy
formation. Nonetheless, for $\masneu$ values compatible with present
observational constraints, these changes are of the same order of
magnitude as the variance between the predictions of different SAMs
applied to the same \lcdm realisation and due to the different
modelling of the relevant physical processes.

Stronger constraints on the cosmological models are indeed accessible,
should detailed information on the overall DM field be available. In
particular, we show that both galaxy bias and the pairwise velocity
distribution are sensitive to the presence of massive neutrinos (the
effects being larger with larger $\masneu$); more interestingly both
diagnostics show deviations in the opposite direction with respect to
Early Dark Energy and/or $f(r)$-gravity models. Tests based on these
observables are thus able not only to disentangle runs including
massive neutrinos from \lcdm, but also to discriminate between
different non-\lcdm cosmologies. On the other hand, since a neutrino
background is expected to be present in all cosmological models based
on the Big Bang theory, we expect it to weaken any signal coming from
an Early Dark Energy or $f(r)$-gravity cosmology \citep[see
  e.g.][]{Baldi14}.

Overall, the results presented in this work complement our previous
claims about the effect of cosmological models which deviates from a
standard \lcdm model in the ``Dark sector'', given different
assumptions on the nature and properties of Dark Energy or Dark
Matter. Such analysis confirms the relevance of studying the
modifications induced in galaxy properties by alternative cosmologies,
in order to tailor effective cosmological tests to be performed with
galaxy surveys. Forthcoming planned space missions like EUCLID
\citep{Laureijs11} are indeed designed to describe the large scale
structure, using both weak lensing and slitless spectroscopy
techniques, and compare it with the spatial distribution of
galaxies. In this framework, it is crucial to build mock catalogues
covering as many cosmologies as possible: in a forthcoming work, we
plan to further extend this approach by considering other cosmological
models like the coupled DE scenarios \citep[see
  e.g.][]{Baldi12}. Finally, in this series of papers we consider
cosmological volumes best suited to study the galaxy mass function
over a wide range of stellar and halo masses (i.e. from $\sim 10^9
\msun$ to $\sim 10^{12} \msun$): we plan to extend the analysis using
larger box-size simulations to improve the statistical power of our
approach, especially at large scales (i.e. galaxy clusters).

\section*{Acknowledgements}
We thank Carmelita Carbone for enlightening discussions. FF
acknowledges financial contribution from the grants PRIN MIUR 2009
``The Intergalactic Medium as a probe of the growth of cosmic
structures'' and PRIN INAF 2010 ``From the dawn of galaxy formation''.
FVN and MV are supported by the ERC Starting Grant ``cosmoIGM'' and
partially supported by INFN IS PD51 "INDARK".  We acknowledge partial
support from "Consorzio per la Fisica - Trieste".

\bibliographystyle{mn2e}
\bibliography{fontanot}

\end{document}